\documentclass[aps,prb,preprint,superscriptaddress,citeautoscript,showpacs]{revtex4-1}

\usepackage{graphicx}
\usepackage{amsmath}
\usepackage{hyperref}
\usepackage{epstopdf}
\usepackage{url}

\begin{document}

%\preprint{}

\title{Excitonic coupling dominates the homogeneous photoluminescence excitation linewidth in semicrystalline polymeric semiconductors}



\author{Pascal~Gr\'egoire}
%\thanks{}
\author{Eleonora~Vella} 
\affiliation{D\'epartement de physique \& Regroupement qu\'eb\'ecois sur les mat\'eriaux de pointe, Universit\'e de Montr\'eal, C.P.\ 6128, Succursale centre-ville, Montr\'eal H3C~3J7, Canada}
\author{Matthew~Dyson}
\affiliation{Department of Physics, Imperial College London, South Kensington Campus, London SW7~2AZ, United~Kingdom}
\affiliation{Department of Materials, Imperial College London, South Kensington Campus, London SW7~2AZ, United~Kingdom}
\author{Claudia~M.~Baz\'an}
\affiliation{D\'epartement de physique \& Regroupement qu\'eb\'ecois sur les mat\'eriaux de pointe, Universit\'e de Montr\'eal, C.P.\ 6128, Succursale centre-ville, Montr\'eal H3C~3J7, Canada}
\author{Richard~Leonelli}
\affiliation{D\'epartement de physique \& Regroupement qu\'eb\'ecois sur les mat\'eriaux de pointe, Universit\'e de Montr\'eal, C.P.\ 6128, Succursale centre-ville, Montr\'eal H3C~3J7, Canada}
\author{Natalie~Stingelin}
\affiliation{Department of Materials, Imperial College London, South Kensington Campus, London SW7~2AZ, United~Kingdom}
\affiliation{School of Materials Science and Engineering and School of Chemical and Biochemical Engineering, Georgia Institute of Technology, Atlanta, Georgia 30332-0245, USA}
\author{Paul~N.~Stavrinou}
\affiliation{Department of Engineering Science, University of Oxford, Parks Road, Oxford, OX1~3PJ, United~Kingdom}
\affiliation{D\'epartement de physique \& Regroupement qu\'eb\'ecois sur les mat\'eriaux de pointe, Universit\'e de Montr\'eal, C.P.\ 6128, Succursale centre-ville, Montr\'eal H3C~3J7, Canada}
\author{Eric~R.~Bittner}
\affiliation{Departments of Chemistry and Physics, University of Houston, Houston, Texas 77204-5003, USA}
\author{Carlos~Silva}
\email[e-mail: ]{carlos.silva@umontreal.ca}
\affiliation{D\'epartement de physique \& Regroupement qu\'eb\'ecois sur les mat\'eriaux de pointe, Universit\'e de Montr\'eal, C.P.\ 6128, Succursale centre-ville, Montr\'eal H3C~3J7, Canada}
\affiliation{Department of Physics, Imperial College London, South Kensington Campus, London SW7~2AZ, United~Kingdom}

\date{\today}

\begin{abstract}
We measure and model the homogeneous excitation linewidth of regioregular poly(3-hexylthiophene), a model semicrystalline polymeric semiconductor,
by means of two-dimensional coherent photoluminescence excitation spectroscopy. At a temperature of 8\,K, we extract a linewidth of $\sim 90$\,meV full-width-at-half-maximum, which is a significant fraction of the total linewidth. We interpret this homogeneous broadening as a consequence of interchain exciton coupling and discuss it within the context of a weakly coupled aggregate model.
\end{abstract}

%\keywords{}

\maketitle

Unravelling optical and electronic properties from disordered energy landscapes, and correlating these to complex solid-state microstructures, are fundamental issues in the materials science of disordered semiconductors. 
For example, in polymeric semiconductors~\cite{Treat2015,Ostroverkhova2016}, the structure-property interdependence is such that the excitation spectral lineshapes are governed fundamentally by the interplay of electronic interactions occurring within a given polymer chain and between adjacent chain segments, which is a strong function of microstructure~\cite{Spano2014}. Regioregular poly(3-hexylthiophene) (P3HT) is amongst several materials that have provided a platform to develop understanding of such interplay, since its absorption and photoluminescence (PL) spectral lineshape can be analyzed using well-established concepts pertaining to molecular aggregates. This allows to extract intricate information on the magnitude of excitonic coupling, the extent of energetic disorder and the amount of correlation in the disordered energy landscape~\cite{Yamagata2012,Paquin2013}. However, extensive PL measurements on isolated P3HT chains at low temperature reveal that the energy of the PL spectrum can vary over a large spectral range throughout the visible~\cite{Thiessen2013}. This suggests instead that bulk lineshapes are predominantly the product of a large inhomogeneously broadened distribution of chromophore energies, which is conformational in origin at the single-molecule level, and which add up to produce the bulk spectrum. Indeed, mixed quantum-classical atomistic simulations demonstrate that energetic disorder of single P3HT chains arises through the interplay of excited-state delocalization and electron-hole polarization, governed by intrachain torsional disorder, and giving rise to a heterogeneous distribution of chromophore energies~\cite{Simine2017}.  
Intriguingly, while the PL origin emission spanned a broad spectral range at 4\,K, its linewidth only varied from $\sim10$--30\,meV and the spectral lineshape was essentially invariant~\cite{Thiessen2013}, in contrast to the broad lineshapes with very strong microstructure dependence in bulk films~\cite{Paquin2013}. The single-chain PL data at low temperature reported in ref.~\citenum{Thiessen2013} thus pose the following fundamental materials physics question: are excitonic models based on photophysical aggregates sufficient to describe quantitatively spectral lineshapes in semicrystalline polymeric semiconductors, or is the governing factor the conformation-induced disorder of individual chains?  The answer to this question necessitates knowledge of the homogeneous spectral linewidth of P3HT in the solid state, with polymer molecular weight and dispersity characteristics comparable to the material reported in ref.~\citenum{Thiessen2013}. A rigorous measurement of the homogeneous lineshape of bulk P3HT films has not been reported to the best of our knowledge.

To address this question, here we evaluate the homogeneous excitation linewidth in a P3HT bulk film in order to compare with the linewidths measured using single-molecule spectroscopic techniques. We do so by means of two-dimensional coherent PL excitation (2D-PLE) spectroscopy~\cite{Tekavec2007}. This ultrafast probe belongs to the family of 2D optical spectroscopies~\cite{Fuller2015} that measure a nonlinear optical response via the time-integrated and spectrally-integrated PL intensity. We choose to implement 2D coherent spectroscopy because these techniques can distinguish between homogeneous and inhomogeneous linewidths~\cite{Tokmakoff2000}. We find that, at 8\,K, the solid-state homogeneous linewidth is up to an order of magnitude broader than in isolated chains. We interpret these observations as excitonic coupling effects in photophysical aggregates~\cite{Spano2014}.

P3HT with weight-average molecular weight ($M_{\textrm{w}}$) of 113\,kg/mol, dispersity (D) of 1.5, and 99\% regioregularity was cast from a chloroform solution as described in detail in Supplemental Material~\footnote{See Supplemental Material at [URL will be inserted by publisher] for details of the experimental measurements and analysis.}. In this high $M_{\textrm{w}}$ regime, the microstructure of P3HT films solution-cast from good solvents consists of crystalline domains embedded in chain-entangled phases~\cite{Reid2012}, with exciton coherence length of $\sim5$ thiophene monomer units along chains, and extending over $\sim2$ chain segments in the interchain direction under the processing conditions employed here~\cite{Paquin2013}.

% FIG 1 ABSORPTION AND 2D PLE SCHEMATIC%%%%%%%%%%%%%%%%%%%%%%%%%%%%%%%%%%
\begin{figure}
\includegraphics[width=7.75cm]{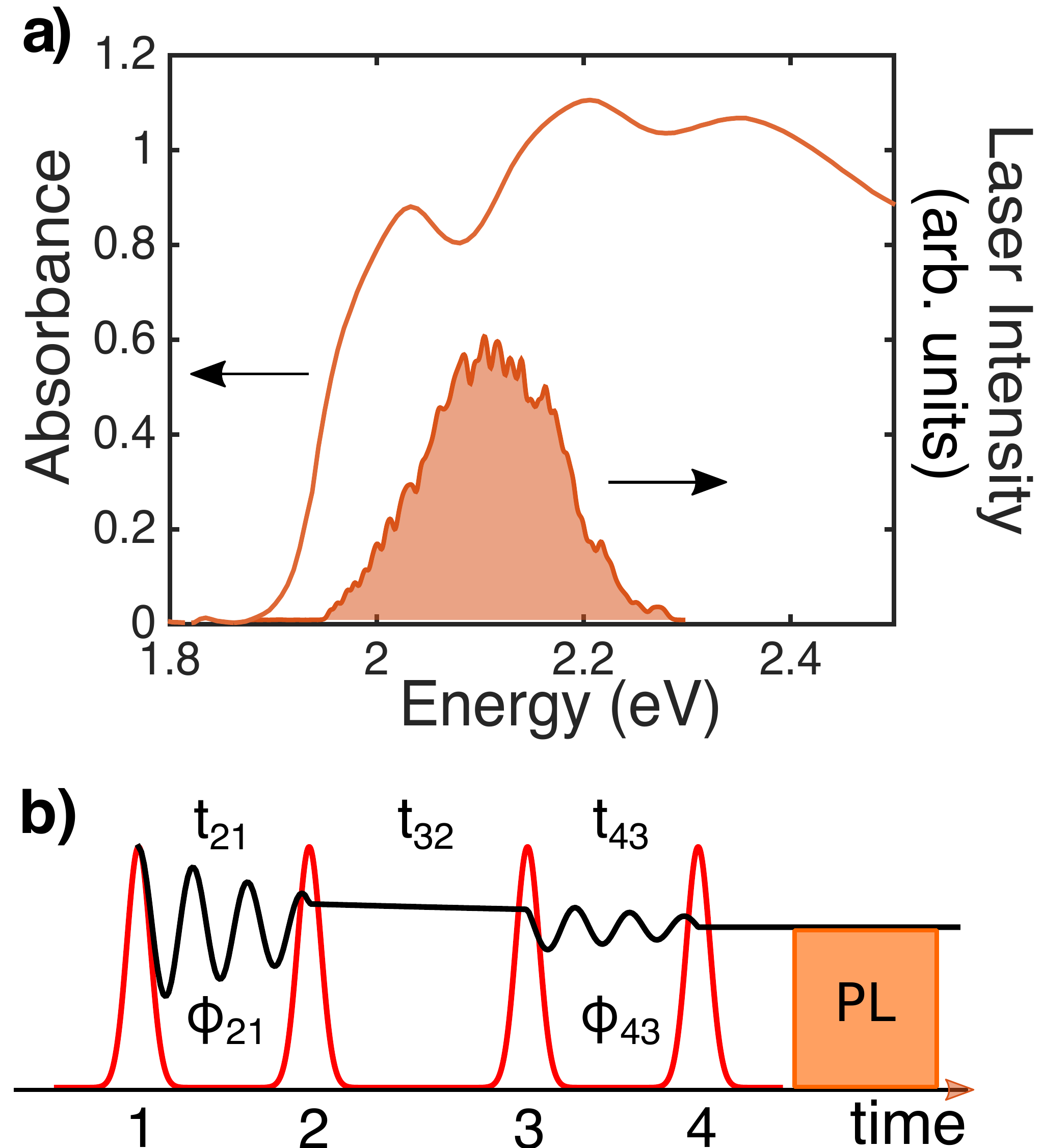}
\caption{(color online). (a) Absorption spectrum of P3HT measured at 10\,K and the spectrum of the femtosecond laser used in this work. (b) Schematic of the pulse sequence for the 2D-PLE measurements~\cite{Note1}. We used transform-limited, 13-fs pulses. The spectrally and temporally integrated PL intensity is measured as a function of the time delays $t_{21}$ and $t_{43}$ at a fixed population time $t_{32}$. The phase differences $\phi_{21}$ and $\phi_{43}$ are modulated to isolate the desired nonlinear signals. \label{fig1}}
\end{figure}
%%%%%%%%%%%%%%%%%%%%%%%%%%%%%%%%%%%%%%%%%%%%%%%%%%%%%%%%%%%%

Fig.~\ref{fig1}(a) shows the absorption spectrum of a film produced by this procedure. Within the weakly-coupled H-aggregate model developed by Spano~\cite{Spano2005,*Spano2007}, we extract a free-exciton bandwidth of $W=63 \pm 5$\,meV from the ratio of the 0--0 and 0--1 absorbance peaks~\cite{Clark2007,Clark2009}. This is consistent with $W$ reported by Paquin et~al.~\cite{Paquin2013} for films of P3HT with similar $M_{\textrm{w}}$ and processed under similar conditions as the sample studied here.  

We examine the excitation lineshape by means of 2D-PLE spectroscopy. The collinear pulse sequence used to acquire a 2D spectrum is shown in Fig.~\ref{fig1}(b)~\cite{Note1}. We collect the time-integrated PL intensity as a function of inter-pulse delays $t_{21}$ and $t_{43}$ (coherence times) at a fixed population time delay $t_{32}$. Acousto-optic modulators modulate the phase difference of the first pulse pair $\phi_{21}$ (pulses 1 and 2, which we denote as the `pump' pulses) and the second pulse pair $\phi_{43}$ (pulses 3 and 4, `probe' pulses) at distinct frequencies in the kHz range. This allows demodulation of the measured PL intensity using phase-sensitive detection with reference waveforms built from a mix of these two phase modulation waveforms. We thus extract simultaneously the so-called \emph{rephasing} and \emph{nonrephasing} signals at each delay point $t_{21}$ and $t_{43}$, producing 2D coherence decay maps, which are Fourier-transformed on both time axes to obtain the 2D coherent excitation spectrum. A detailed description of the experimental technique and analysis is provided in Supplemental Material~\cite{Note1}. 
All measurements reported here were taken using a pulse fluence of $4.2$\,$\mu$J\,cm$^{-2}$ and at a temperature of 8\,K. 
 
%% FIG 2 ABSORPTIVE SPECTRA %%%%%%%%%%%%%%%%%%%%%%%
\begin{figure*}
\includegraphics[width=16cm]{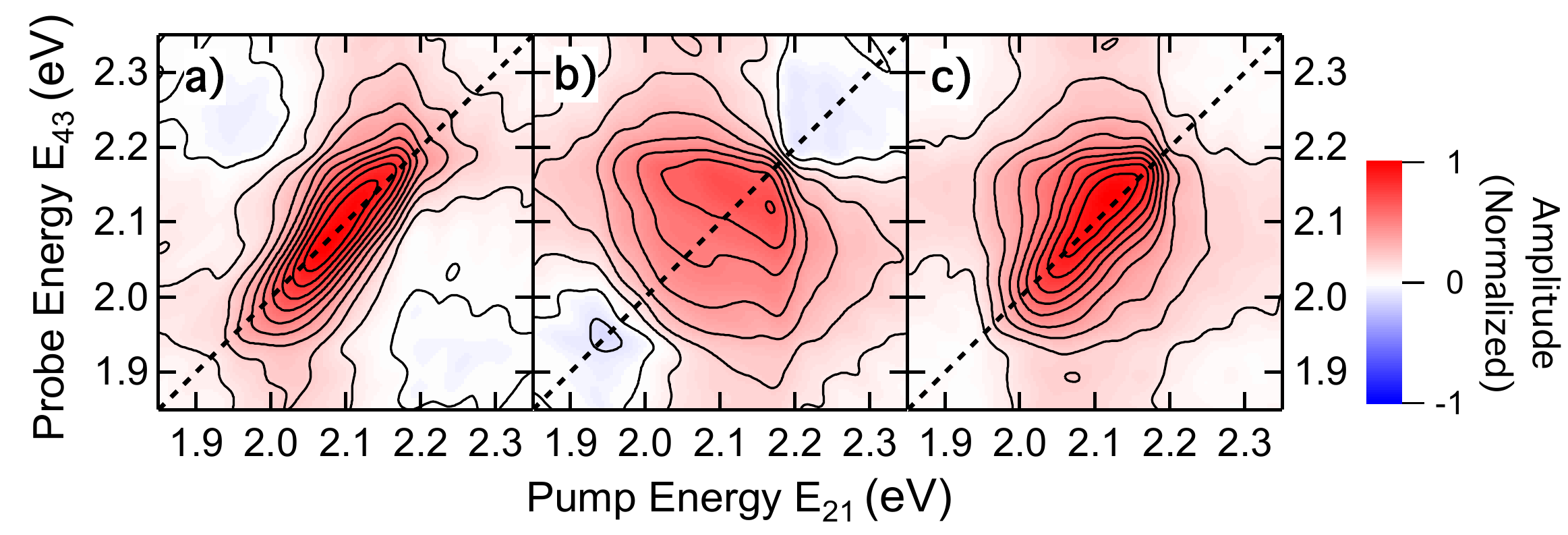}
\caption{(color online). Real part of (a) rephasing, (b) nonrephasing and (c) total correlation spectra, which correspond to the purely absorptive spectral component, measured at 8\,K. These data were acquired with a population waiting time $t_{32} = 50$\,fs. \label{figabs}}
\end{figure*}
%%%%%%%%%%%%%%%%%%%%%%%%%%%%%%%%%%%%%%%%%%

A typical set of early-time 2D-PLE spectra is depicted in Fig.~\ref{figabs}. The purely absorptive (real-part) spectrum (Fig.~\ref{figabs}(c)) is the sum of the rephasing (Fig.~\ref{figabs}(a)) and the nonrephasing (Fig.~\ref{figabs}(b)) spectral components. The complete response, including the imaginary spectra, is presented in Supplemental Material~\cite{Note1}. We observe in all these a diagonal peak at 2.05\,eV, corresponding to the 0--0 absorption feature in Fig.~\ref{fig1}(a). 
To higher energies along the diagonal, we observe the low-energy tail of the 0--1 replica. We also observe weak 0--0/0--1 cross peaks, most prominent at $(E_{21}, E_{43}) = (2.05, 2.19)$\,eV in the nonrephasing spectrum. These data are consistent with previous measurements in P3HT films~\cite{DeSio2016,Song2015} and in P3HT aggregates suspended in solution~\cite{Song2014}. Slight differences in lineshape can be ascribed to differences in microstructure and in our distinct experimental implementation of the 2D coherent probe~\cite{Perdomo-Ortiz2012}.

% FIG 3 simulations %%%%%%%%%%%%%%%%%%%%%%%%
\begin{figure}[]%htbp
 \includegraphics[width=7.75cm]{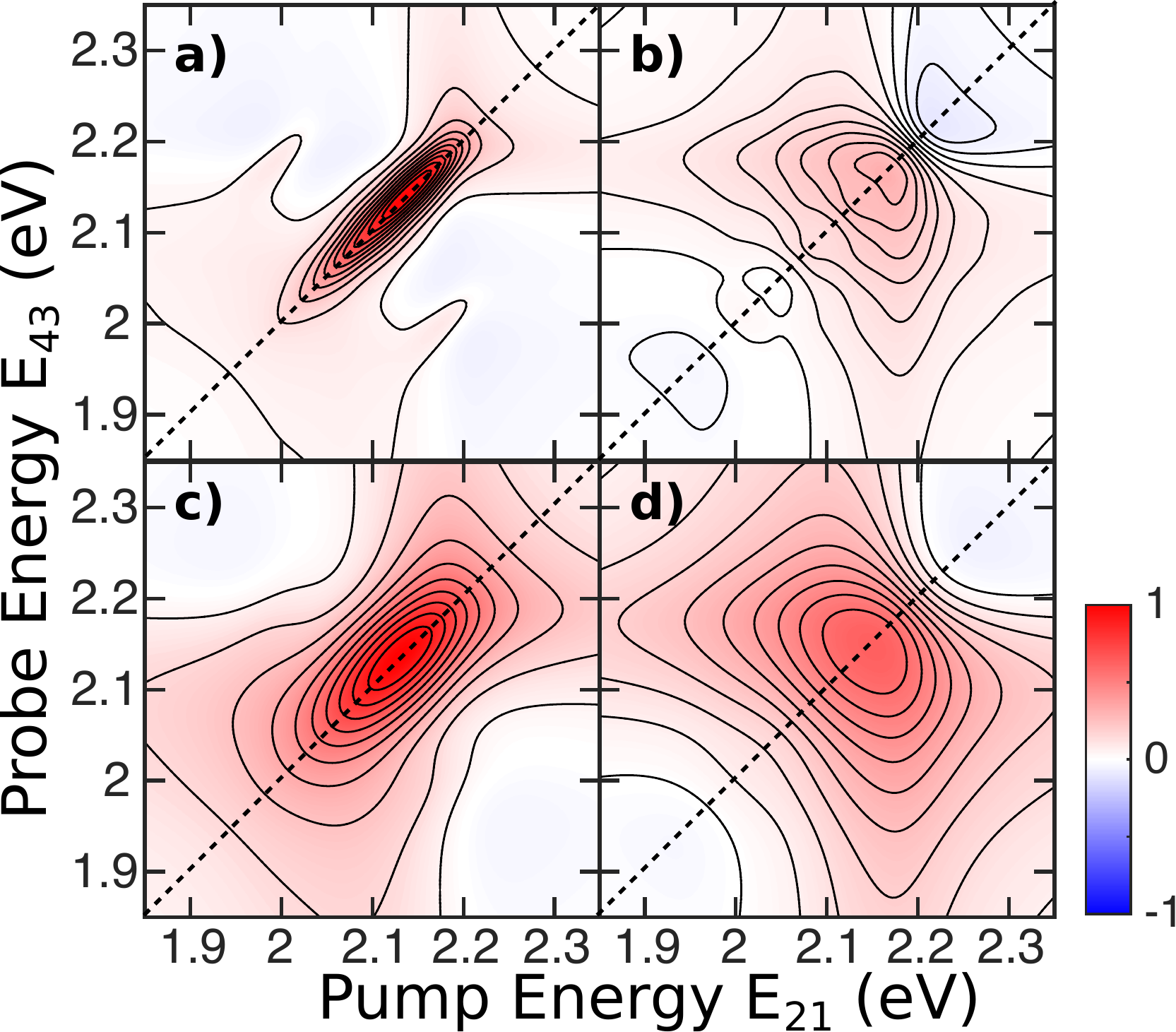}
\caption{(color online). Numerical simulations of the P3HT rephasing (left) and nonrephasing response (right). The real part is displayed using  $2\gamma=30$\,meV (top) and $2\gamma=90$\,meV (bottom), both with inhomogeneous broadening $\sigma=130$\,meV. Here $\gamma$ is the dephasing rate and $\sigma$ is the inhomogeneous spectral width.
 \label{fig_sim}}
\end{figure}
%%%%%%%%%%%%%%%%%%%%%%%%%%%%%

In general, the diagonal width of the zero-time 2D coherent rephasing spectrum represents the inhomogeneous linewidth, here limited by the excitation laser bandwidth, and the antidiagonal spectrum displays the homogeneous dephasing~\cite{Tokmakoff2000}. However, as pointed out by Siemens et~al.~\cite{Siemens2010}, inhomogeneous and homogeneous broadening mechanisms need to be evaluated simultaneously to extract reliable quantitative values from 2D coherent excitation spectra, which is challenging to do analytically for systems with cross peaks. We have carried out numerical simulations of 2D coherent spectra using a three-state model, consisting of the ground state and 0--0 and 0--1 vibronic replica, as described in Supplemental Material~\cite{Note1}.  We have  taken into account the 0--0 and 0--1 relative spectral weight and the excitation laser spectrum explicitly. We also include a phenomenological Gaussian inhomogeneous broadening with spectral width $\sigma$ for the 0--0 and 0--1 peaks, and a common dephasing rate $\gamma$. In the simulations displayed in Fig.~\ref{fig_sim}, we have used $\sigma=130$\,meV full-width-at-half-maximum, which is consistent with the disorder widths used in to model absorption and PL lineshapes in P3HT~\cite{Paquin2013}. Fig.~\ref{fig_sim}(a) and (b) show the real part of the rephasing and nonrephasing spectrum, respectively, for $2\gamma=30$\,meV, which corresponds to the upper end of the spectral linewidth distribution measured in the single-chain experiments of Thiessen et~al.~\cite{Thiessen2013} In Fig.~\ref{fig_sim}(c) and (d), we show the simulated spectra with $2\gamma=90$\,meV. In both cases, 0--0/0--1 cross peaks can be identified in the simulated spectra, as observed experimentally in Fig.~\ref{figabs}. However, in the situation of narrow homogeneous broadening (top row), these peaks are more readily identified compared to when the homogeneous linewidth is a more significant fraction of the inhomogeneous linewidth (bottom row). We find that we can best represent our measured 2D-PLE spectra using $2\gamma=90$\,meV, indicating that the homogeneous linewidth in the solid state is substantially broader than the spectral linewidths measured for isolated polymer chains.

% FIG 4 2D PHOTON echo %%%%%%%%%%%%%%%%%%%%%%%%
\begin{figure}
\includegraphics[width=7.75cm]{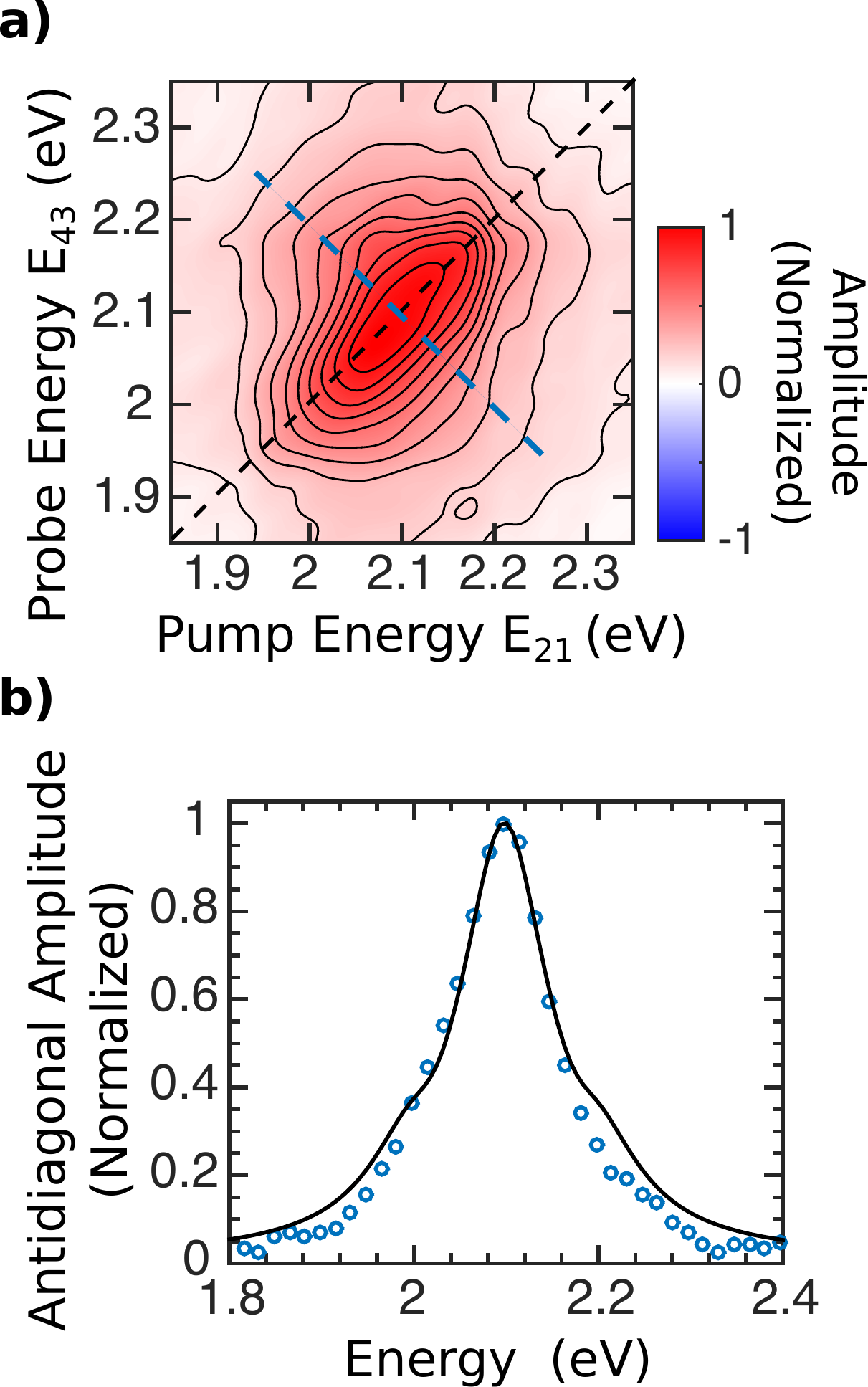}
\caption{(color online). (a) Modulus of the measured 2D-PLE rephasing spectrum displayed in Fig.~\ref{figabs}(a). The elongated feature along the diagonal is characteristic of inhomogeneous broadening, limited by the excitation laser spectrum, while the antidiagonal slice (blue dashed line) reveals the homogeneously-broadened spectrum~\cite{Tokmakoff2000}.  (b) Antidiagonal slice of the photon echo signal at 2.10\,eV (blue open circles) corresponding to the dashed blue line in part (a). The black continuous curve displays the corresponding slice for the simulated spectra using $2\gamma=90$\,meV and $\sigma=130$\,meV (shown in Fig~\ref{fig_sim}(c)).\label{fig_reph}}
\end{figure}
%%%%%%%%%%%%%%%%%%%%%%%%%%%%%
 
Returning to the experimental measurement, Fig.~\ref{fig_reph}(a) displays the modulus of the 2D-PLE rephasing spectrum, i.e.\ the photon-echo spectrum, in this neat P3HT film at 8\,K. We have chosen to analyze this spectrum measured at a  population time $t_{32}= 50$\,fs to avoid artefacts due to time-overlap of the $\sim13$-fs pulse pairs. We demonstrate in Supplemental Material that the spectral linewidth does not evolve over this initial population waiting period~\cite{Note1}, thus this measurement is equivalent to the usual zero-time photon-echo spectrum. 
We display the antidiagonal slice, centered at $E_{21}=E_{43}=2.10$\,eV, over the blue dashed line in Fig.~\ref{fig_reph}(a), as open circles in Fig.~\ref{fig_reph}(b). The corresponding slice for our simulation with $2\gamma=90$\,meV  is represented by the black continuous curve, which reproduces the width of the diagonal peak on top of the 0--0/0--1 cross peaks. From this comparison, we estimate that the homogeneous linewidth in our P3HT films is $90\pm10$\,meV. 
The spectral resolution of this time-domain measurement is, in principle, limited by the measurement range in $t_{21}$ and $t_{43}$, the coherence time variables depicted in Fig.~\ref{fig1}(b). For the spectrum reported in Fig.~\ref{fig_reph}(a), we have measured 2D coherence decay functions over 100-fs time windows~\cite{Note1}, which implies a  resolution of $\sim 20$\,meV. The homogeneous linewidth extracted in Fig.~\ref{fig_reph}(b) is therefore well within our experimental resolution.

This PL excitation linewidth is roughly an order of magnitude broader than the lower limit in the single-chain PL spectra reported by Thiessen et~al.~\cite{Thiessen2013}, who investigated P3HT with $M_{\textrm{w}}$ that is near the transition in polymer microstructure from a nonentangled, paraffinic to an entangled, semicrystalline, two-phase morphology in bulk films~\cite{Paquin2013,Reid2012}. In that work, while the emission energy of individual P3HT chains varied substantially over much of the visible range, the PL spectral lineshape  was found to be nearly invariant over measurement of hundreds of different chains. The lineshape was found to be strongly dominated by the 0--0 emission peak, with a narrow width spanning from $\sim10$ to $25$\,meV in the spectral range accessible by our 2D-PLE measurements. While PL excitation (PLE) spectral measurements on single molecules are significantly more challenging than PL experiments, we expect PLE spectra of comparable width. We must therefore invoke a homogeneous broadening mechanism that accounts for the large difference in linewidth in going from isolated P3HT chains to that in the bulk.

The homogeneous excitation linewidth $2 \gamma$  is related to the optical dephasing time $T_2$ via the time-energy uncertainty principle, $T_2 = \hbar/\gamma$. From the bulk spectrum of Fig.~\ref{fig_reph}(b), we extract $T_2 = 15 \pm 2$\,fs. In  an isolated chain, we infer from the measurements of Thiessen et~al.~\cite{Thiessen2013} a value of $T_2 \sim 50$--130\,fs. Employing two-color, three-pulse photon echo peak shift measurements  in dilute chloroform solution, Wells and Blank concluded that intrachain excitons in P3HT reflect dynamic broadening due to strong, selective exciton-vibrational coupling that is dominated by only two motions --- one high-frequency bond stretch and a low-frequency torsional motion~\cite{Wells2008}. Although at the low concentrations at which those measurements were carried out chain aggregation was considered to be suppressed, Raithenel et~al.\ point out that even in solution at such low concentrations, `loose' aggregates are formed, and in fact can dominate the optical response~\cite{Raithel2016}. Therefore, given that the rephasing measurements reported by Wells and Blank were carried out in solution and at room temperature, we might expect that the dephasing dynamics are intermediate between those measured here and those inferred from measurements reported by Thiessen et~al. We thus consider that the $T_2$ derived from the low-temperature single-chain measurements reflect a limiting case of that intrachain dephasing mechanism given the low temperature of the measurement and that the polymer chains were on a silica substrate under vacuum. Qualitatively, we suggest that the additional broadening seen in our bulk measurements results from interchain photophysical aggregate effects.  

To examine this idea more formally, let us consider a simple two-dimensional 
free-exciton model of an aggregate composed of polymer 
chains of length $L_{x}$ that are assembled to 
form a lamellar stack of length $L_{y}$. 
Excitonic coupling effects, both along and between chains, 
are incorporated  by an intrachain hopping integral $t$, 
and an interchain hopping integral $J$.  
 Typically, one finds that $J/t \sim 0.1$ 
within a hybrid HJ aggregate model~\cite{Yamagata2012,Paquin2013}. 
For an isolated chain, the homogeneous linewidth scales as~\cite{Note1}
\begin{eqnarray}
\gamma_{iso} \propto \frac{t}{L_x^2}.
\label{eq:iso}
\end{eqnarray}
This is consistent with the results presented by Arias et~al.\ on J aggregates~\cite{Arias2013}.
Including the interchain term and assuming $L_x \gg L_y$, one obtains a similar 
expression along the interchain stacking direction~\cite{Note1},
\begin{eqnarray}
\gamma_{agg} \propto \frac{J}{L_y^2}.
\label{eq:agg}
\end{eqnarray}
The ratio of our measured homogeneous linewidth to the most narrow linewidths measured by Thiessen et~al.~\cite{Thiessen2013} allows us to relate $J/t$ 
to the aspect ratio of the aggregate domain: 
\begin{eqnarray}
r =\frac{\gamma_{iso}}{\gamma_{agg}}  =  \frac{t}{J}\left(\frac{L_y}{L_x}\right)^2 \sim 0.09.
\end{eqnarray}
Rearranging this, and assuming $J/t \sim 0.1$, we find that 
\begin{eqnarray}
\frac{L_y}{L_x} \propto \sqrt{r\frac{J}{t}}\sim 0.09.
\label{eq:aspect}
\end{eqnarray}
This aspect ratio is consistent with two-dimensional exciton coherence functions in P3HT films, modeled by Paquin et~al.\ using a disordered Holstein lattice Hamiltonian representing hybrid HJ aggregates~\cite{Paquin2013}.  
This is particularly so considering that our picture here does not include electron-vibration coupling effects accounted for in the Holstein lattice, which would localize the exciton further within the polymer backbone. Furthermore, the coherence function modeled by Paquin et~al.\ was based on the steady-state PL spectral lineshape, which reflects an equilibrium distribution of more highly localized excitons. 
Therefore, our simple model rationalizes the nearly order-of-magnitude difference in dephasing times in the bulk compared to isolated chains. 
Physically, additional fluctuations in the exciton energy due to fluctuations in excitonic coupling between $\pi$ electrons across chains gives rise to more rapid dephasing than in isolated chains, even in the weak coupling aggregate regime.
Indeed, Yamagata et~al.\ highlight the role of short-range charge-transfer-mediated coupling in small-molecule aggregates~\cite{Yamagata2014}, and we suggest that these types of intermolecular interactions are responsible for the additional optical dephasing pathways observed in our measurements reported here. 
These arguments support our proposal that large homogeneous broadening in the bulk, in contrast to isolated chain, is due to interchain photophysical aggregate effects.

These solid-state contributions on the homogeneous linewidth are consistent with other photophysical consequences of even weak aggregate formation, revealed by means of single-aggregate spectroscopy measurements. For example, Raithenel et~al.\ show that subtle aggregation of P3HT chains produces strong red shifts and broadening in comparison to spectra from  isolated-chain~\cite{Raithel2016}.
Hu et~al.\ observed that long-range exciton transport depends strongly on aggregate order~\cite{Hu2014}, further highlighting the importance of interchain interactions on exciton dynamics. In a further study, Hu et~al.\ demonstrated that dielectric-induced stabilization of nonradiative charge-transfer states very sensitively controls PL quantum efficiencies in suspended P3HT aggregates~\cite{Hu2015}, consistent with the role proposed by Yamagata et~al.\ of charge-transfer components of excitonic coupling~\cite{Yamagata2014}, indicating once again the key role the environment play in dictating photophysics in aggregates.

The clear excitonic signatures on the homogeneous PL excitation linewidth, derived from our 2D-PLE measurements, reveal the importance of employing photophysical aggregate models to describe the bulk optical properties of polymeric semiconductors. This conclusion is in contrast to that reached by Thiessen et~al.~\cite{Thiessen2013}, where the bulk spectral lineshapes were assumed dominated by the chain-conformation-dependent combination of single chromophores within a broad inhomogeneous distribution. We put forth here that excitonic effects are not minor contributions to the optical properties of this class of semiconductor materials, and we highlight the richness of information on exciton-environment coupling dynamics reported by the nonlinear spectroscopic techniques employed in this work.

\begin{acknowledgments}
CS, PG, and EV are indebted to Prof.\ Andy Marcus and Dr.\ Julia Widom for their enthusiastic assistance in the construction of the experimental setup. CS acknowledges funding from the Engineering and Physical Sciences Research Council of Canada (NSERC), the Canada Foundation for Innovation,  
and the Universit\'e de Montr\'eal University Research Chair. PG acknowledges funding from NSERC via a Doctoral Postgraduate Scholarship. 
The work at the University of Houston was funded in part by the National Science Foundation (CHE-1362006)  and the Robert A. Welch Foundation (E-1337). 
Furthermore, MD, NS and PS thank the United Kingdom's Engineering and Physical Sciences Research Council (EPSRC) for funding via the Centre for Doctoral Training in Plastic Electronics Materials (EP/G037515/1). 
M.D, N.S and P.N.S gratefully acknowledge the U.K. Engineering and Physical Sciences Research Council (EP/ G037515/1).
\end{acknowledgments}

%\bibliography{library}

\end{document}